\documentclass[twocolumn,showpacs]{revtex4}
\usepackage{psfig}

\begin{document}
\draft

\title{\hfill{\tiny FZJ--IKP(TH)--2004--12} \\[1.8em]
  Comment on 'Photoproduction of $\eta$--Mesic ${^3}$He'}

\author{C. Hanhart}

\address{IKP(TH), Forschungszentrum J\"ulich, D--52428 J\"ulich, Germany}

\begin{abstract} 
In a recent paper by the TAPS collaboration \cite{exp} a first measurement of
a bound system of an $\eta$ meson and a ${^3}$He nucleus was reported.
In this comment we critically reexamine the interpretation of
the data and  show that the data prefers a solution where there is
no bound state present. Given the low statistics of the measurement,
however,
it does not exclude the existence of a bound state.
\end{abstract}


\maketitle

\vspace{0.8cm}
\newcommand{\boldpi}{\mbox{\boldmath $\pi$}}
\newcommand{\boldtau}{\mbox{\boldmath $\tau$}}
\newcommand{\boldT}{\mbox{\boldmath $T$}}
\newcommand{\gaprox}{$ {\raisebox{-.6ex}{{$\stackrel{\textstyle >}{\sim}$}}} $}
\newcommand{\saprox}{$ {\raisebox{-.6ex}{{$\stackrel{\textstyle <}{\sim}$}}} $}

The interaction of $\eta$ mesons with nucleons is strong and attractive due
mainly to the presence of the $S_{11}(1535)$ resonance that strongly couples
to this system. Consequently it is expected that the $\eta$ meson should be
bound in sufficiently heavy nuclei. So far, however, it is unclear what mass
number is sufficient. Some authors predicted a bound state to occur on nuclei
as light as ${^3}$He~\cite{Wycech,Belyaev1,Belyaev2,Rakityansky,Fix1,Wilkin1},
whereas others expect binding only for heavier nuclei
\cite{Haider1,Liu,Chiang,Haider2}. Until recently no direct experimental
evidence for the existence of $\eta$--mesic nuclei was available. Only the
presence of a strong $\eta$--nucleus interaction was seen experimentally in
strong final state interaction effects in reactions like $pn\to \eta
d$\cite{Calen}, $pd\to \eta {^3}$He\cite{Berger,Mayer}, and $dd\to \eta
{^4}$He\cite{Frascaria}.

Thus it was a big step forward from the experimental side when this year the
TAPS collaboration reported positive evidence for $\eta$--mesic ${^3}$He.
Besides a strong deviation in the angular shape of $\gamma {^3}\mbox{He}\to
\eta {^3}$He from the expectation for quasi--free production
(the cross section is flat instead of forward peaked), a structure
was observed in the cross section $\gamma {^3}\mbox{He}\to
\pi^0pX$ just below the $\eta$ production threshold. These signatures were
taken as strong evidence for the existence of $\eta$--mesic ${^3}$He.

It should be clear that the former evidence---a flat $\eta {^3}$He angular
distribution in the close--to--threshold regime is a hint solely for a strong
$s$--wave $\eta {^3}$He interaction that leads to a relative suppression of the
impulse term with respect to the $s$--wave multiple scattering terms. Thus,
given what we already know about the strong $\eta {^3}$He interaction,
 a flat angular
distribution in the close--to--threshold regime should be expected. A closer
look at the structure in the cross section $\gamma {^3}\mbox{He}\to
\pi^0pX$ is the focus of this comment.

\begin{figure}[t]
\vspace*{-0mm}\hspace*{-3mm}
\psfig{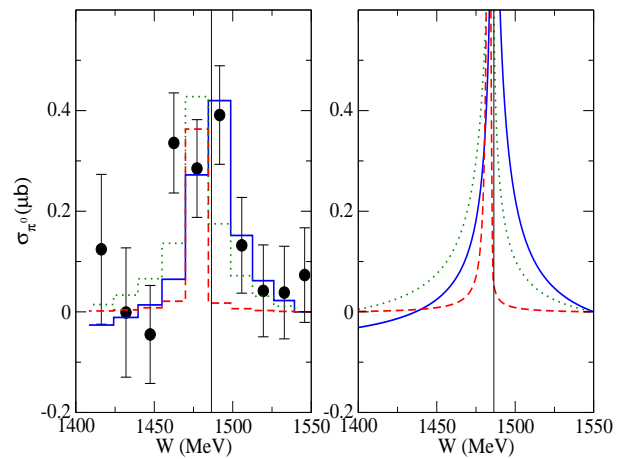}\vspace*{-2mm}
\caption{Comparison of the various fits to the data, as a function of the
  reduced photon energy $W$ defined in Ref. \protect\cite{exp}. Shown are the
  results in the absence of a background interference ($B=0$ in Eq.
  (\protect\ref{full})). The left panel corresponds to the calculation using
  the same binning as the data, whereas the right panel shows the results
  with no binning. The solid line corresponds to $a=(+4,1)$ fm, the dashed
  one to $a=(-4,1)$ fm and the dotted one to $a=(0,3.5)$ fm.
The vertical line at $W=1486.4$ MeV indicates the  position of the $\eta {^3}$He threshold.}
\label{resonly} 
\end{figure}

The structure reported by the TAPS collaboration was fitted with a
Breit--Wigner function. In its non--relativistic form the scattering
amplitude then is
\begin{equation}
f_{BW} \propto \left(E-E_R+\frac i2 \Gamma\right)^{-1} \ ,
\label{BW}
\end{equation} 
where $\Gamma$ is assumed to be constant.
The parameters deduced were $(4.4{\pm}4.1)$ MeV and
$(25.6{\pm6.1})$ MeV for the binding energy and the width, respectively.
However, since the position of the signal
 coincides with the $\eta$
production threshold, one might wonder whether this is more a cusp than
the signal of a bound state. 

Already in 1976, when studying the light scalar mesons $a_0$ and $f_0$,
Flatt{\'e} observed that in the presence of thresholds the Breit--Wigner form
of Eq.  (\ref{BW}) is to be modified to include the momentum--dependence of
the elastic width \cite{flatte} (for a more recent discussion of threshold
effects in various system we refer to Ref. \cite{Bugg}). Thus Eq. (\ref{BW})
should be changed according to
$$
\Gamma \to \Gamma_{inel}+\Gamma_{el} \ ,
$$
where $\Gamma_{inel}$ and $\Gamma_{el}$ denote the inelastic and the elastic
width (in our case with respect to the $\eta {^3}$He channel)
 of the resonance respectively. Here $\Gamma_{inel}$ can be assumed constant; however, 
$\Gamma_{el}$ has to vanish at the elastic threshold! Thus, for an $s$--wave
structure, one gets 
$$
\Gamma_{el} = g_{eff}k \ ,
$$
where $k$ denotes the momentum of the $\eta$ relative to the ${^3}$He nucleus
and, above the production threshold, may be written as $k=\sqrt{2\mu E}$. Here
$\mu$ denotes the reduced mass of the $\eta$He system and $E$ is its kinetic
energy. In the region below threshold, however, $k=i\sqrt{-2\mu E}$.
Thus, we find that if a structure is predominantly inelastic, a Breit--Wigner
might still be a good approximation, even in the proximity of a threshold;
however, if a structure is predominantly elastic, using a Breit--Wigner is not
justified. 

A dynamically generated singularity, like a bound state, also dominates
the final state interaction in the $\eta$He system. In addition,
 if a production reaction is short--ranged (typical
momentum transfer significantly larger than any other scale of the problem)
the final state interaction is universal (independent of the reaction) and can
be related to the elastic scattering of the outgoing particles~\cite{wam}
which reads in the effective range approximation
\begin{equation}
f_{sc} \propto \left(1/a+rk^2/2-ik\right)^{-1} \ .
\label{era}
\end{equation}
   Recently
the world data set on the reaction $pd\to\eta {^3}$He was analyzed
\cite{unserhe}. This study led to quite constrained values for the real and
imaginary part of the $\eta {^3}$He scattering length, namely
\begin{equation}
 a=(\pm 4.3 \pm
0.3 \, , \ 0.5\pm 0.5) \ \mbox{fm} \ ,
\label{scl}
\end{equation}
where the first number refers to the real part and the second number to the
imaginary part---in the analysis the effective range term of Eq. (\ref{era})
was neglected ($r=0$). Note, these numbers
where found from a fit to the world data set. However, this dataset is
inconsistent and if we use only the newest data in the fit the scattering
length is less constrained; see Ref.  \cite{unserhe} for details. We come back
to this point below.
The
two signs given in front of the real part indicate that $\eta$ production
data can not fix the sign of the real part. A positive sign would point at a
virtual state (a singularity on the unphysical sheet), whereas a negative sign
would point at the existence of a bound state.
 In Ref. \cite{mixing} it was
stressed that isospin--violating ratios of pion production cross sections taken
in the vicinity of the $\eta$ production threshold should be a good tool to
fix the sign of the real part. 
 It is
important to understand whether or not the TAPS measurement is sufficient to
decide on the sign of the
real part of the scattering length. 

The Flatt{\'e} form discussed above can be easily matched to the effective
range approximation of Eq. (\ref{era}) \cite{evidence}.
One thus finds that neglecting the effective range term in  Eq. (\ref{era})
is equivalent to assume
that $g_{eff}$ is sufficiently large that in the region of interest $E$ can be
neglected in Eq. (\ref{BW}). 
In Ref. \cite{evidence} it was argued that this should be a good
approximation if the structure of interest is dynamically generated and the
singularity is close to the threshold, as is the case here. In addition, the
role played by the effective range term in the $\eta {^3}$He final state
interaction is completely unclear and the data for $pd\to \eta {^3}$He could
be very well fitted using $r=0$.

\begin{figure}[t]
\vspace*{-0mm}\hspace*{-3mm}
\psfig{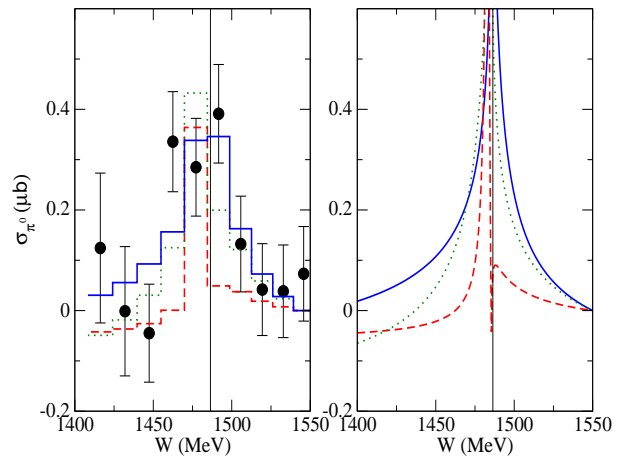}\vspace*{-2mm}
\caption{Comparison of the various fits to the data, as a function of the
  reduced photon energy $W$ defined in Ref. \protect\cite{exp}. Shown are the results 
for a dominance of the interference with the background
 ($B$ large in Eq. (\protect\ref{full})). The
  left panel corresponds to the calculation using the same binning as the
  data, whereas the right panel shows the results with no binning. The
  solid line corresponds to $a=(+4,1)$ fm, the dashed one to $a=(-4,1)$ fm and
  the dotted one to $a=(0,3.5)$ fm.
The vertical line at $W=1486.4$ MeV indicates the position of the $\eta {^3}$He threshold.}
\label{backonly} 
\end{figure}

There is one additional comment necessary before
we can  apply Eq. (\ref{era}) to the TAPS data: there is in principle some
interference with the background. Thus, what was identified as the resonance
signal might well have some contribution from an interference term, and the
full signal my be written as
\begin{equation}
N\left(2\mbox{Re}(Bf^{res})+\left|f^{res}\right|^2\right) \ ,
\label{full}
\end{equation}
where $B$ is some complex number parameterizing that part of the background
that is allowed to interfere with the resonance signal and $N$ is a measure of
the total strength of the signal. Therefore, we performed three different
fits: fit 1 included only the pure resonance signal ($B=0$; only $N$ as a free
parameter); fit 2 included only the interference term ($B\to \infty$; $N$ and
the phase of $B$ as a free parameter); and fit 3 considered the full structure
(thus here we have 3 free parameters: $N$, $|B|$ and the phase of $B$). As it
turned out, the $\chi ^2$ per degree of freedom for the two scenarios
(positive and negative real part of the scattering length) was almost the same
in all three cases and thus for illustration in Figs. \ref{resonly} and
\ref{backonly} we only show the results of the first two fits, where the left
panel corresponds to the results after binning in accordance with
that of
the experiment and the right panel corresponds to the unbinned results.  To
keep the numbers of free parameters low we choose $a=(\pm 4,1)$ fm.  In both
figures the dashed line corresponds to a negative real part (indicating
 the existence of a bound state) and the solid line corresponds
to a positive real part (indicating a virtual state). The fit gave a
$\chi ^2$ per degree of freedom of 1 for the latter case, whereas it was worse
than 3 in the former.  Thus the data prefers the solution that corresponds to
a virtual state, although the existence of a bound state can not be excluded,
given the quality of the data.
Note, already in Ref. \cite{shnm} the interpretation of the TAPS data as a
bound state was questioned.

There is one important comment to be added: the data set for $pd\to \eta
{^3}$He shows some inconsistencies. As discussed in detail in Ref.
\cite{unserhe}, a fit to just the most recent data allows for a significantly
broader band of scattering lengths: then even the case of a vanishing real
part is not excluded (together with Im$(a)$=3.5 fm). To illustrate the impact
of this scattering length in Figs. \ref{resonly} and \ref{backonly} we also
show the corresponding results as the dotted curve. As can be
seen, this fit is almost equally good as that with the positive scattering
length ($\chi ^2$ per degree of freedom of about 2).

Thus we conclude that the data on $\gamma {^3}\mbox{He}\to \pi p X$ recently
measured by the TAPS collaboration does not allow for a conclusion on the
existence of a bound system of $\eta$ and ${^3}$He. To improve the situation
the measurement should be redone with improved statistics to allow for 
smaller energy bins. In addition, to permit
 an unambiguous interpretation of  $\gamma {^3}\mbox{He}\to \pi p
X$,  more refined information on the $\eta {^3}$He scattering length is
needed. Fortunately, this will be available soon from measurements performed
at COSY \cite{COSY1,COSY2,COSY3}. The present paper clearly shows the usefulness of
a combined analysis of data from both electromagnetic and hadronic probes.

{\bf Acknowledgement}

We thank B. Krusche, J.A. Niskanen, V. Metag, M. Pfeiffer,  and C. Wilkin for useful discussions and
J. Durso for numerous editorial remarks.

\end{document}